# Towards Security as a Service to Protect the Critical Resources of Mobile Computing Devices

Abdulrahman Alreshidi

College of Computer Science and Engineering
University of Ha'il
Ha'il, Saudi Arabia

*Abstract*—**Mobile computing is fast replacing the traditional computing paradigms by offering its users to exploit portable computations and context-aware communications. Despite the benefits of mobile computing, such as portability and context-sensitivity, there are some critical challenges, such as resource poverty of mobile devices and security of mobile user's data that must be addressed. Implementing the security mechanisms to execute on mobile devices can be challenging as mobile devices lack the required processor, memory and battery resources to support continuous and long-term execution of computation intensive tasks. Cloud computing model can provide virtually unlimited hardware, software, and service resources to compensate for the resource poverty of mobile devices. In recent years, there is a lot of research and development of solutions and frameworks that preserve the security and privacy of mobile devices and their data. However, there has been little effort to secure mobile devices while also supporting an efficient utilization of the limited resources available on mobile devices. In this paper, we propose Security as a Service for mobile devices (SeaaS for mobile) that integrates mobile computing and cloud computing technologies to secure the critical resources of mobile devices. The proposed solution aims to support 1) security for the data critical resources of mobile devices, and 2) security as a service by cloud servers for an efficient utilization of the mobile device resources. We demonstrate the security as a service based on a practical scenario for the security of mobile devices. The evaluation results show that the proposed solution is 1) accurate to detect the potential security threats, and is 2) computationally efficient for mobile devices. The proposed solution as part of ongoing research provides the foundations to develop a framework to address SeaaS for mobile. The proposed solution aims to advance the research state-of-the-art on software engineering, mobile cloud computing, while it specifically focuses exploiting cloud-based services to secure mobile devices.**

## I. INTRODUCTION

Mobile computing has fast emerged as a pervasive technology that empowers its users to exploit portability and context-awareness to perform a variety of tasks on the go [1]. Specifically, mobile computing can utilize the embedded sensors (i.e., GPS, Accelerometer as hardware resources) that can be combined with freely available apps (i.e., location services, maps as software resources) of a mobile device to support context-aware and portable computing [2]. For example, the mobile users can enable GPS based location sensing to get live updates about traffic conditions or recommendations about the places/events of interest based on their geographical proximity. Despite these benefits, mobile computing in general and mobile devices in particular face two primary challenges [8], [9]. The first challenge relates to the resource poverty, i.e., the availability of limited processing, memory and battery resources to a mobile device. The second challenge is to protect the integrity of the mobile device that is prone to the threats of data security and privacy in a context-aware environment. The security threats relate to the security critical resources that are hardware (e.g., Microphone, GPS sensor) and software (e.g., Contacts List, Photos) resources. For example, a third part game installed on a mobile device can try to maliciously access the Microphone to spy on user's voice conversation or look into user's contact list for information [3]. If such private information can be compromised or exploited by entities with malicious access, it can put user's information and device's data on security risk [14].

### A. Research Challenges

There is a need for a rigorous security mechanism(s) that protects the data critical resources of a mobile device to support secure mobile computing. However, any rigorous security solution(s) that continuously execute on a mobile device to protect itself may be impractical mainly due to the computation, memory and battery specific resource poverty of the mobile devices. There is a need for solutions that must ensure a rigorous security as well as efficient resource utilization of a mobile device [4]. Cloud commuting represents an opportunistic computing model that relies on the 'pay-per-use' hardware and software services that can be used and released as required. Cloud computing model offers three main types of services referred to as Software as a Service (SaaS), Platform as a Service (PaaS), and Infrastructure as a Service (IaaS) [5], [15], [16]. In recent years, mobile and cloud computing technologies have been unified to enable Mobile Cloud Computing (MCC) as state-of-the-art mobile computing technology. Specifically, in MCC a mobile device represents a portable and context-aware user interface (as front-end technology) that relies on the resource sufficient cloud-based servers to perform the complex computations (as back-end technology) - off-loaded by mobile devices to the cloud-based servers. For example, the solutions of MCC in [6] allow mobile devices to off-load their computation intensive tasks to cloud-based servers to prolong the battery life and enhance the processor and memory performance of the mobile devices.

This means that existing solutions of MCC have been successful to support mobile devices that are portable, context-sensitive, as well as resource sufficient by relying on cloud computing.

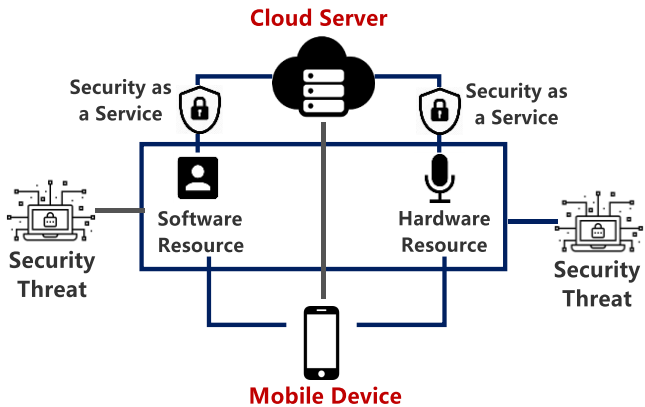

Fig. 1. Overview of the Solution to Support SeaaS for Mobile Devices.

### B. Proposed Solution, Contributions, and Assumptions

We propose to address the security related issues that are relevant to the critical resources of mobile devices by integrating the mobile computing and cloud computing technologies. In the proposed solution, we aim to develop a framework that provides Security as a Service for mobile devices (SeaaS for mobile) offered as a cloud-based service to secure mobile computing. We define SeaaS for mobile as 'security policies that are executed on cloud-based servers to continuously monitor and protect the critical (hardware and software) resources of a mobile device'. Critical resources of a mobile device are any hardware and software resources that produce or consume the information that can be prone to security threats and malicious access. We provide an overview of the proposed solution in Fig. 1. The figure highlights that the mobile device and cloud servers are connected. The cloud server runs the SeaaS to continuously monitor the critical hardware (e.g., Microphone) as well as software (e.g., Contacts) resource. Security specific issues of the mobile device resources are offloaded to the cloud-based server. Any potentially malicious access is detected and is eliminated or minimized by the cloud-based SeaaS. The primary contributions of research are summarized below.

- *Unification of Mobile and Cloud Computing Technologies* to exploit security as a service to protect mobile devices that operates in ad-hoc (unsecured) network environment.

- *Off-loading the Execution of Security Mechanism* and policies on cloud-based servers to support secure and efficient mobile computing.

Based on the proposed contributions, we have the following assumptions for the proposed solution.

- *A Continuous Network Connectivity* is required between the mobile device and cloud-server for the monitoring and protection of the device resources by cloud-based services.

- *Integration of Mobile and Cloud Computing* where a mobile device represents a client, whereas the cloud-based server(s) act as security provision resources.

The rest of the paper is organized as follows. Background details about Mobile Computing Environment are provided in Section 2. Related Work is presented in Section 3. The Proposed Solution is presented in Section 4. Solution Demonstration and Evaluation Results are presented in Section 5. Conclusions and Future Research are detailed in Section 6.

## II. BACKGROUND ABOUT MOBILE COMPUTING ENVIRONMENT

We now present an overview of the mobile computing environment as in Fig. 2. We briefly discuss the elements of the mobile computing environment in the context of mobile security. The concepts and terms introduced here will be used in the remainder of this paper. As highlighted in Fig. 2, the elements of a mobile computing environment are introduced below. In Fig. 2, we have adopted this general model of mobile computing environment from [14].

### A. Mobile Device

It represents any (handheld) equipment or machine that allows its user to perform computation, in-formation sharing and other activities in a mobility-driven environment. A mobile device is a combination of hard-ware (e.g. GPS sensors, Camera) that is manipulated by means of software apps (e.g. Location Tracker, Image Editor) and both are vulnerable to the security threats.

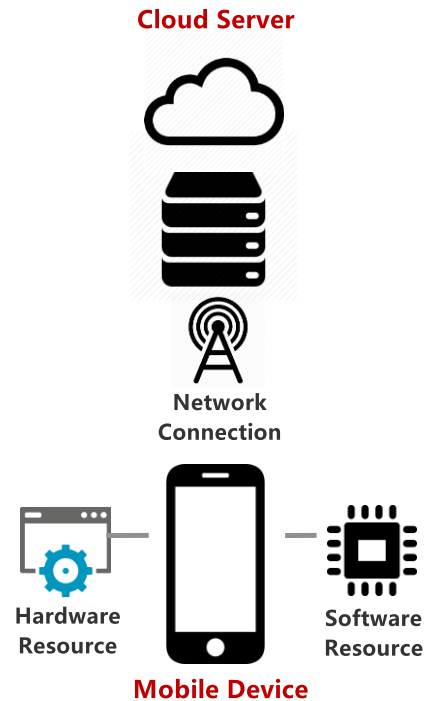

Fig. 2. Overview of the mobile computing environment.

- **Hardware Resources** are physical parts of a device such as sensors, processors that are sources of private

information. For example, to uniquely identify a device's hardware, the device-specific information of a mobile device is represented with Unique Device Identifier (UDID) or International Mobile Equipment Identity (IMEI) that are subject to security threats. In terms of software, a mobile device equipped with Global Position System (GPS) represents a scenario where user's location or context can be revealed or leaked to location sensing services.

- **Software Resources** are the applications or data that contains the useful information or private data of the device users. For example, the Contacts List or Photos represent the software resources of a mobile device and these resources are at a risk of comprising the private information.

### B. Mobile Server

It provides an infrastructure that allows a mobile device (client) to store or retrieve data, request the desired services and to off-load the data for computation on the server. In Fig. 1, the server can be of many types (e.g., Communication, Proxy, and Database) that can provide a lot of new functionality such as location services, look-ups in directory services and enabling distributed data storage and processing. With an emergence of the mobile-cloud computing a mobile device as a resource-constrained computer can exploit virtually unlimited computation and storage via resource sufficient cloud server. Specifically, cloud servers have been proven successful solutions to compensate the limited computation, storage and power re-sources of mobile devices. In a mobile-cloud computing environment, the mobile acts as a context-aware and port-able client to capture and display data. In comparison, the cloud represents a backend server that supports the computation and storage of all the data off-loaded to it by a mobile device. A mobile server is prone to invasion that can compromise the security of mobile data residing on the server.

### C. Network Connectivity

It allows a mobile device to communicate with a server through network connections such as Wi-Fi or Bluetooth signals. This means that in addition to the data in a mobile device or the one residing on the server, the communication channel can be attacked to compromise the data that travels between a mobile and its corresponding server. A typical example is the attack on location queries that travel between a mobile device and location-providing server [3]. In addition, a direct communication between mobile devices is also subject to security threats on the communication channel. Based on the illustration in Fig. 1, we conclude that security in mobile computing environment allows protection and preservation of (user and de-vice) data or information deemed as private from acts of malice [5], [6].

### III. RELATED RESEARCH ON SECURE MOBILE COMPUTING

In this section, first we highlight the existing research (Section 3.A) and then discuss some proposed solutions as tools and frameworks (Section 3.B) that enable or enhance mobile computing security. By presenting the most relevant related work, we justify the scope and proposed contributions of our research.

It is vital to mention that, according to the GSMA real-time tracker, the world is currently home to more than 7.2 billion mobile device connections, where 0 to 7 billion connections have been achieved in just three decades [7]. This implies that mobile connections are currently growing about five times faster than the human population. In a recent report, the Homeland Security has highlighted that security specific threats to mobile computing such as user location tracking, banking and transactions fraud, ransom-ware, identity theft puts at risk not just mobile device users, but the mobile carriers as well as infrastructure providers [3].

### A. Summary of Existing Research on Security for Mobile Computing Environments

A survey-based study highlights the potential threats and proposed solutions for devices that operate in mobile and ad-hoc networks [8]. In recent years, there is a lot of focus on solutions to enable or enhance the security of the mobile devices. Specifically, in [9], the authors have highlighted the potential and a huge market for android applications, however; there are concerns relating to the security and privacy issues of these apps. Unless there is a strong mechanism for a device to protect itself, the widely available apps are subject to potential security threats. Moreover, a mobile device has limited resources, i.e.; memory and processing power which means that it becomes challenging to maintain security of its data.

The study [10], suggests a balance between IT infrastructure overhead and system security. The study has considered four application level security systems and evaluated them against a pre-defined scheme i.e., systems support for critical security related services such as authentication, authorization, maintainability, re-usability, productivity etc. On the basis of the evaluation results, the study concludes that that none of the selected systems fulfilled the evaluation schemes. In the mobile computing context, the challenge lies in providing a robust security mechanism while also supporting the efficiency of the devices memory, computation and energy efficiency.

TABLE I. A COMPARISON SUMMARY OF THE EXISTING FRAMEWORKS FOR MOBILE COMPUTING SECURITY.

| Proposed Solution | Mobile Computing | Cloud Computing | Proposed Contributions |
|---|---|---|---|
| Xposed Framework | ✓ | | Mobile Apps Monitoring |
| Android Monkey UI Exerciser | | ✓ | Testing Android Apps |
| Resource Description Framework (RDF) | ✓ | | Encode Data for Exchange |
| Proposed Solution | ✓ | ✓ | Mobile Device Security |

## B. Summary of Frameworks and Tools for Mobile Computing Security

After presenting the research challenges, we also highlight some existing frameworks and tool support that aim to automate and enhance adaptive security solutions. These tools and frameworks are mostly proof of the research concepts.

- *Xposed* framework is an open-source framework that allows monitoring of the installed apps can change system settings i.e., behaviour of the system or the installed apps as per the security needs [11]. The users can change the settings of the framework at runtime and can also bring the changes based on those dynamic settings.

- *Android Monkey UI Exerciser* framework is basically a third-party tool which helps in testing android applications. It is basically a command line tool which works with adb tool (Android Debug Bridge) [12]. It is basically used to perform stress testing on the android applications and to report back the errors, if occurred.

- *Resource Description Framework (RDF)* is the framework for data interchanges over the internet. This framework is capable of encoding, exchanging and reusing the structured metadata [13]. It is an application of XML that provides a description to specify the resources and the associated threats.

### 1) Comparative Summary – Existing vs Proposed Research

A comparative summary between the existing solutions and the proposed solution is provided in Table I. Specifically, Table I highlights that in comparison to the existing Solutions, the proposed solution exploits mobile-cloud computing technologies to enable the security of mobile devices. Based on the discussion above, there is a lack of research that supports cloud-based security as a service to protect resource constrained mobile devices. The proposed research aims to off-load security mechanisms to cloud-based servers to support secure and efficient mobile computing.

## IV. A LAYERED SOLUTION FOR SECURITY AS A SERVICE FOR MOBILE COMPUTING

We now present an overview of the framework in Fig. 3. Fig. 3 highlights that the proposed solution is based on a layered architecture system. Specifically, the proposed solution has the following two layers namely Mobile Computing Layer and Cloud Computing Layer.

### A. Layer I - Security Critical Mobile Computing Layer

It is the front-end layers that allow a user to perform computing and communication in a portable and context-aware fashion. As highlighted in Fig. 3, the mobile computing layer has some hardware and software resources that are critical from a security point of view. For example, the *Microphone* that is a hardware resource must be protected from any malicious access to comprise the privacy of *user's voice* and *audio messaging*. In a similar scenario from Fig. 3, *Contacts* that represents a software resource containing the critical information such as contacts' *names*, *numbers*, *email*, *photos* that can be compromised. The intent for any unauthorized access to the hardware and software resources can be due to spying on user's private information or selling user's private information to the third part advertisement providers. In such circumstances there is a need to secure mobile computing resources.

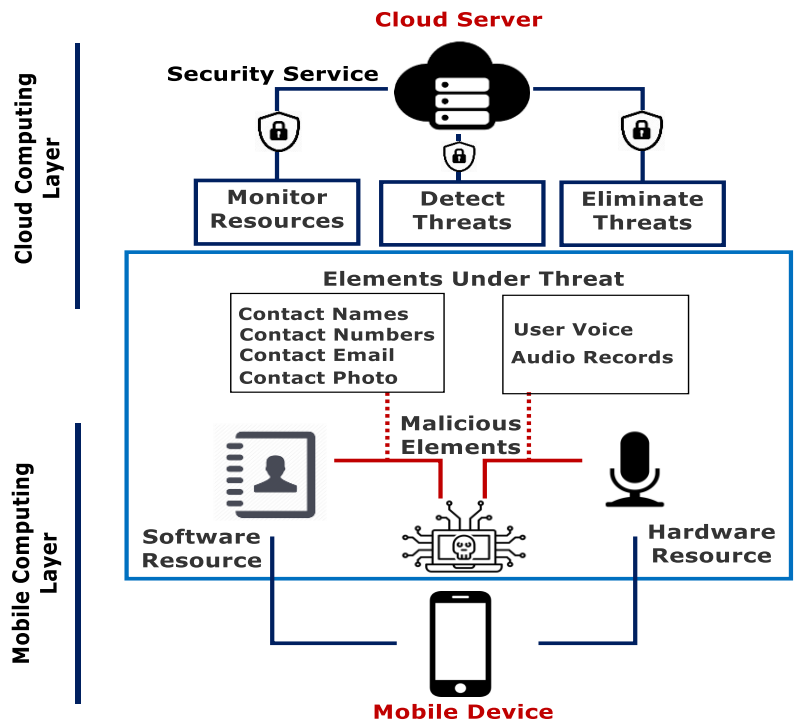

Fig. 3. Overview of the mobile cloud based security as a service.

However, the main challenge relates to the resource poverty of mobile devices that cannot support an effective execution of the rigorous and computational intensive security protocols.

### B. Layer I - Security Critical Mobile Computing Layer

To compensate for the resource constrained mobile devices, the backend cloud computing layer can provide the security as a service for the critical resources of a mobile device [16]. As highlighted in Fig. 3, the cloud-based Security as a service can relive a mobile device from securing its resources in a way that cloud-based server continuously monitors the critical resources of a device and secures them by eliminating and or minimizing any security threats. As highlighted in Fig. 3, the cloud based Security as a Service follows a three steps process that supports: 1) Monitoring of the device resources, 2) Detection of any unwanted and potentially malicious access as a security threat, and 3) Eliminating or minimizing the security threat to secure resources of a mobile device.

## V. DEMONSTRATION AND EVALUATION OF THE FRAMEWORK

After presenting the overview of the solution, we now demonstrate the usability of the framework based on a case study (Section 5.A). We then present the results of the evaluation of the framework (Section 5.B).

### A. Demonstration of the Security as a Service Framework

The scenarios presented here are taken from [14]. Scenario-based demonstration highlights the applicability of the proposed framework based on scenario-driven approach as in Table II. Table II presents the scenarios for Mobile Computing security.

TABLE II.    OVERVIEW OF SCENARIOS FOR SECURITY AS A SERVICE IN THE CONTEXT OF MOBILE COMPUTING

| **Scenario I -** Malicious Access to Hardware Resources |
|---|
| **Security Challenge**: how to protect a mobile device's hardware resources (e.g.; accelerometer, gyroscope, wireless or GPS sensors) from any act of malice that compromises the de-vice's data and security? |
| **Proposed Solution**: The proposed solutions offer Perceptual Monitoring as a mechanism to monitor, control and customise access to a devices sensor to safeguard user's private data (e.g.; locations, actions, movement). Such perceptual monitors enable or enhance a device's security by working as: <ul><li>Monitor the access of the device's sensors.</li><li>Customise the sensor usage policy of third-party apps (e.g.; grant, deny, or selective permission).</li><li>Runtime modification of the sensor access permission as per user's needs and requirements.</li></ul> |
| **Scenario I -** Malicious Access to Mobile Device Data and Resources |
| **Security Challenge** how to enable the security of mobile devices that prevents or minimizes any malicious access to mobile apps and leak-age of private data? |
| **Proposed Solution** provide **Reconfigurable Security Policies** and device data monitors that are executed on a mobile device. These policies and runtime monitors can dynamically configure their behaviour -depending on the context of data/app being accessed -to prevent or minimise any malicious access to device data and apps. Reconfigurable security policies work as follows: <ul><li>Define access policies for device's data and apps.</li><li>Execute policies as backend processes to monitor malicious access to app or misused data</li><li>Reconfiguration policies as per the context of app or data access and taking into account user's customisation of the policies.</li></ul> |

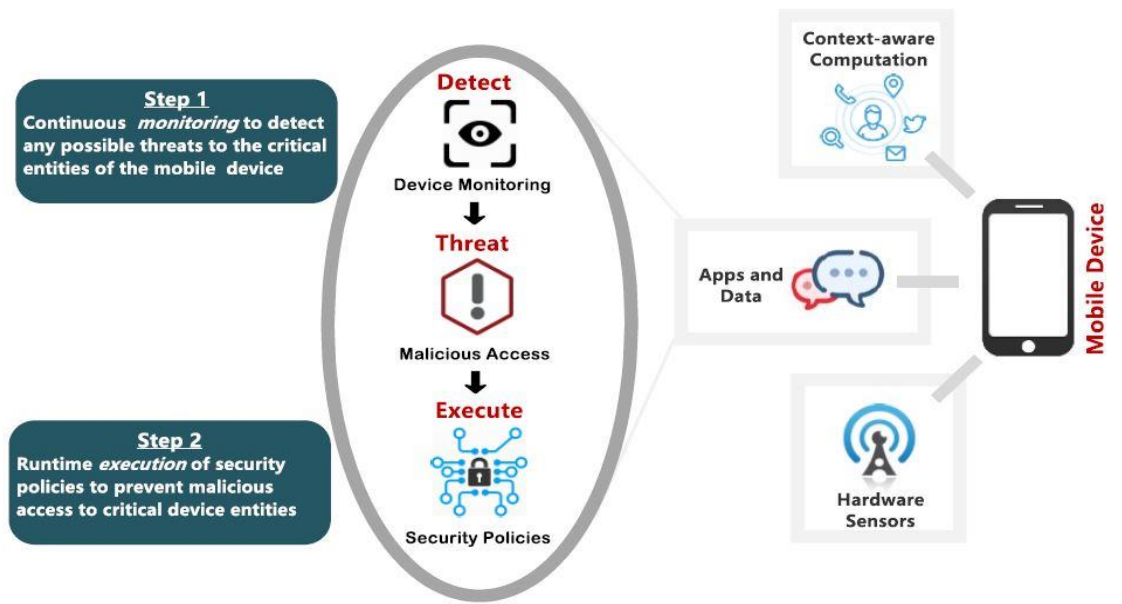

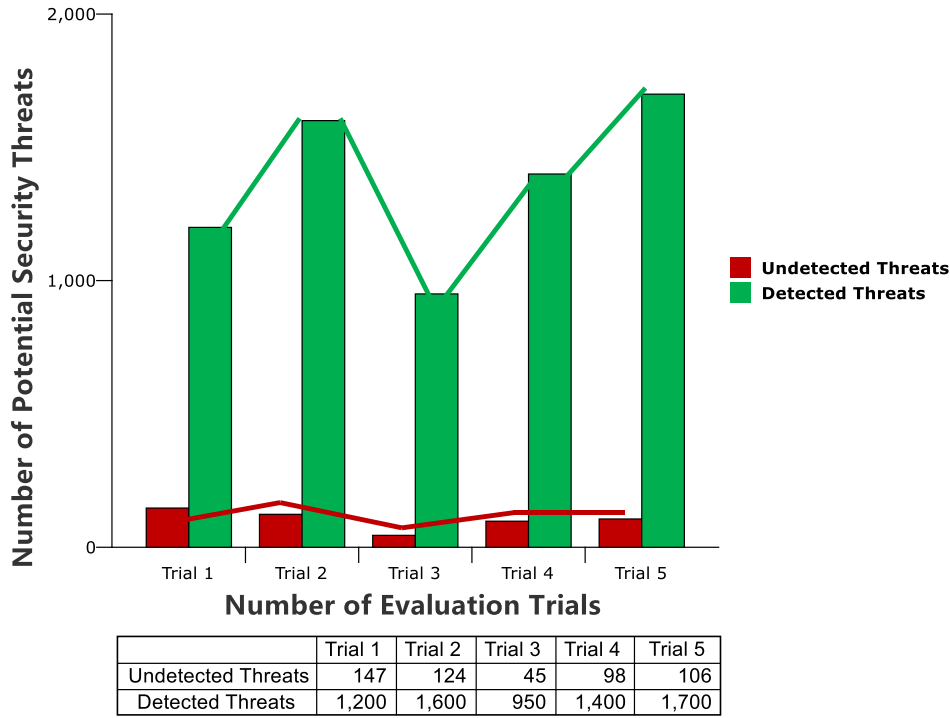

| | Trial 1 | Trial 2 | Trial 3 | Trial 4 | Trial 5 |
|---|---|---|---|---|---|
| Undetected Threats | 147 | 124 | 45 | 98 | 106 |
| Detected Threats | 1,200 | 1,600 | 950 | 1,400 | 1,700 |

Fig. 4. Overview of the mobile cloud based security as a service.

### A. Results of Preliminary Evaluations

After presenting the demonstration of the framework, we now discuss the results of the evaluation to validate the accuracy and efficiency or the formwork. Measurement of both the accuracy and efficiency represent the qualitative evaluation of the framework based on the IS0-IEC-9126 model for software quality [1]. To conduct the preliminary evaluations of the proposed solution, we have used:

- *Mobile Computing Layer:* We have used HTML5 technologies for mobile front-end development to target multiple mobile platforms. Another reason of using hybrid mobile application instead of native mobile application development is that we carry out all performance intensive tasks over cloud layer. Considering the mobile services development via HTML5 technologies, we exploit ionic framework to target android and iOS platform.

- *Cloud Computing Layer:* We exploit Amazon cloud services for storage and computing efficiency. Pricing model adopted by AWS is pay-as-you- go. We launch a virtual server on Amazon cloud called Amazon EC2 instance and set up Red Hat Linux operating system over the instance. We develop server-side application using Node.js and set up Node.js web server on Amazon EC2 Instance. For the sake of efficient data retrieval, we use Mongo DB. We install MongoDB on

Amazon EC2 Instance. To use files and media we utilize Amazon S3 storage services.

- *Accuracy to Detect the Potential Threats:* We now present a summary of the framework's accuracy to detect the potential security threats. An overview of the results of the preliminary evaluations is presented in Fig. 4. Specifically, Fig. 4 shows the total number of trials (X-axis) along with the number of detected/undetected threats. We conducted a total of 5 trials where each trial engaged 10 users on average to evaluate the accuracy of potential security threat detection. As per the ISO/IEC-9126 model, accuracy refers to the system's ability to correctly compute the results. As illustrated in Fig. 4, based on 5 trials, the total number of detected threats is 6850, while the detected threats are 520. Based on the average, the ratio of detected to undetected threats (Detected/Undetected) is 13:1that demonstrates a high-level of accuracy for threat detection.

- *Computational Efficiency of the Proposed Solution:* In addition to the accuracy, as highlighted earlier, we also need a solution that is computationally efficient. By computationally efficient we mean that the proposed solution has an efficient utilisation of the mobile device processor. The efficiency can be enabled by off-loading the computational intensive tasks to the cloud-based servers [6], [14]. Fig. 5 highlights the CPU utilization for the mobile device. Fig. 5 highlights two pieces of information, (i) the ratio of processor consumption and (ii) total number of trials.

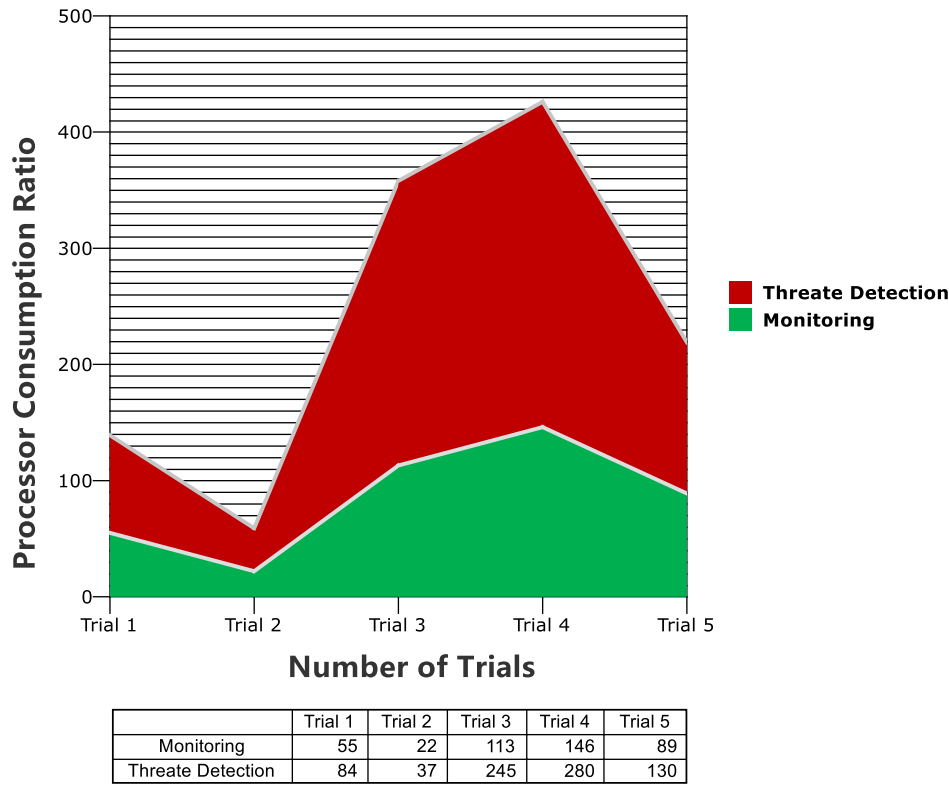

| | Trial 1 | Trial 2 | Trial 3 | Trial 4 | Trial 5 |
|---|---|---|---|---|---|
| Monitoring | 55 | 22 | 113 | 146 | 89 |
| Threate Detection | 84 | 37 | 245 | 280 | 130 |

Fig. 5. Overview of the mobile cloud based security as a service.

The results in Fig. 5 show that while monitoring the potential security threats, how much of the device's processor has been utilised. We conducted a total of 5 trials where each trial engaged 10 users on average to evaluate the accuracy of potential security threat detection. Fig. 5 also demonstrates that only a small percentage of the device's CPU has been used while monitoring for the security threats. Such efficient utilisation of the processor is only possible due to the fact that all computation intensive tasks have been performed on the cloud-based server.

We conclude that in the scope of the proposed solution that is part of the ongoing project, we have only conducted the preliminary evaluation of the proposed solution in terms of system's accuracy and efficiency. The preliminary results show that the proposed solution is efficient and accurate. However, for more concrete and objective evaluation, we need more trials and further development of the system that is part of the future work.

## VI. Conclusions and Future Work

Mobile computing supports portability, context-sensitivity, and enhanced user interaction to replace the traditional computing paradigms. Despite these benefits, mobile computing faces a number of challenges such a resource poverty of a mobile device and threats to the security and privacy of users' information and device's data. Specifically, to address the issues of mobile device security in an efficient way we have proposed s novel solution that relies on the integration of mobile devices and cloud servers to enable or enhance a mobile device's security. The proposed solution aims to address two of the most prominent challenges for mobile computing namely security and efficiency of mobile computing. The proposed solutions exploit a layered approach to address these challenges by offloading the security mechanisms of mobile device to cloud-based servers. Specifically, the front-end layer (i.e., mobile device) represents a portable and context-aware computer that relies on the back-end layer (i.e., cloud server) to monitor and protect the critical resources of the mobile device. The integration of mobile and cloud computing technologies as state of the art mobile computing technology aims to support secure mobile computing.

We have presented the solution architecture and demonstrated its application to enable mobile device security. The results of the preliminary evaluation suggest that proposed solution is (i) accurate for the detection of potential security threats, and (ii) it off-loads computation intensive tasks from mobile devices to cloud-based servers to enable efficient mobile computing. We conclude that the proposed solution can be helpful for

- Advancing the state-of-the-art on mobile computing technology to support mobile-cloud driven security framework.

- Enabling the solution for secure and efficient mobile computing by means of cloud-based security.

*A.  Possible Future Research*

The proposed solution provides a framework and the foundations to develop a comprehensive tool support as a proof-of-the-concept to enable and automate the concept of Security as a Service for the critical resources of mobile devices. Therefore, as part of the future work we mainly focus on the development of the framework that provides an executable solution for further evaluation. Moreover, the proposed solutions need concrete scenarios of security threats that can be executed and analysed to demonstrate the applicability and validation of the proposed solution. We are particularly interested in exploiting the existing algorithms and solutions that can leverage the cloud computing resources to secure mobile devices.